\documentclass[apj]{emulateapj}
\usepackage{graphics,enumerate,amsmath}

\newcommand{\oiw}{\mbox{[\ion{O}{1}] $\lambda$6300}}
\newcommand{\oiiw}{\mbox{[\ion{O}{2}] $\lambda$3727}}

\newcommand{\niibw}{\mbox{[\ion{N}{2}] $\lambda$6583}}

\newcommand{\siiw}{\mbox{[\ion{S}{2}] $\lambda \lambda$6716,6731}}

\newcommand{\oiiibw}{\mbox{[\ion{O}{3}] $\lambda$5007}}

\newcommand{\mgbw}{\hbox{{Mg~b} $\lambda$5170}}
\newcommand{\nadw}{\hbox{{Na~D} $\lambda$5890,5896}}

\newcommand{\nii}{\mbox{[\ion{N}{2}]}}
\newcommand{\hal}{\mbox{H$\alpha$}}
\newcommand{\sii}{\mbox{[\ion{S}{2}]}}

\shorttitle{Precision Spectrophotometry}
\shortauthors{Yan}

\begin{document}
\title{Precision Spectrophotometry at the level of 0.1\%}
\author{Renbin Yan}

\affil{Center for Cosmology and Particle Physics, Department of Physics, New York University, New York, NY 10003, USA; ry9@nyu.edu}

\begin{abstract}
\noindent 
Accurate relative spectrophotometry is critical for many science applications. Small wavelength scale residuals in the flux calibration can significantly impact the measurements of weak emission and absorption features in the spectra.
Using Sloan Digital Sky Survey data, we demonstrate that the average spectra of carefully selected red-sequence galaxies can be used as a spectroscopic standard to improve the relative spectrophotometry precision to 0.1\% on small wavelength scales (from a few to hundreds of Angstroms). We achieve this precision by comparing stacked spectra across tiny redshift intervals. The redshift intervals must be small enough that any systematic stellar population evolution is minimized and less than the spectrophotometric uncertainty. This purely empirical technique does not require any theoretical knowledge of true galaxy spectra. It can be applied to all large spectroscopic galaxy redshift surveys that sample a large number of galaxies in a uniform population.

\rightskip=0pt
\end{abstract}


\section{Motivation}
In astronomical spectroscopy, accurate spectrophotometric calibration is critical for many science applications, but difficult to achieve. All calibrations require a standard source whose intrinsic properties we know. By observing it with the same system (atmosphere, telescope, instrument, detector) as used for observing the science targets, we can infer the throughput of the system as a function of wavelength. However, the accuracy of this calibration is limited by how well we know the standards and their measurement uncertainty. For ground-based astronomical observations, we have to rely on natural standards (but see \citealt{Kaiser08,Kaiser10} and \citealt{Albert11} for attempts to use satellite-mounted standards). Nearly all extragalactic spectroscopy has used stars as the calibration standards \citep[e.g.,][]{Stoughton02}. Therefore, the accuracy is limited by how well we know the true spectra of stars and the uncertainty in the standard star spectra.

The current state-of-the-art spectrophotometric calibration utilized in large surveys can be found in the Sloan Digital Sky Survey (SDSS; \citealt{York00}; \citealt{SDSSDR7}). In SDSS, each plate includes a set of 16 standard stars, which are color-selected to be F8 subdwarfs. Their spectra are reduced in the same way as science targets. The calibration is achieved by comparing the galactic-reddening-corrected spectra of these stars to a grid of theoretical spectra generated from Kurucz model atmospheres. The average ratio of each star to its best-fit model is taken as the flux calibration vector and applied to all the science targets \citep{SDSSDR2}. Finally, the corrected spectra are put on an absolute scale by comparing the synthetic photometry to fiber magnitudes from the SDSS photometry. 

The accuracy of this method is limited by the signal to noise of the standard star spectra and the accuracy of the models. The former limits the confidence in using them to correct for the system response variation on small scales. In SDSS, a fourth-order b-spline with 50 breakpoints is used to smooth the flux calibration vector before applying it to calibrate science targets, thus leaving small-scale variations uncorrected. The theoretical model of the standard stars could also have unknown systematic residuals on small scales. 

\begin{figure*}
\begin{center}
\includegraphics[totalheight=0.35\textheight]{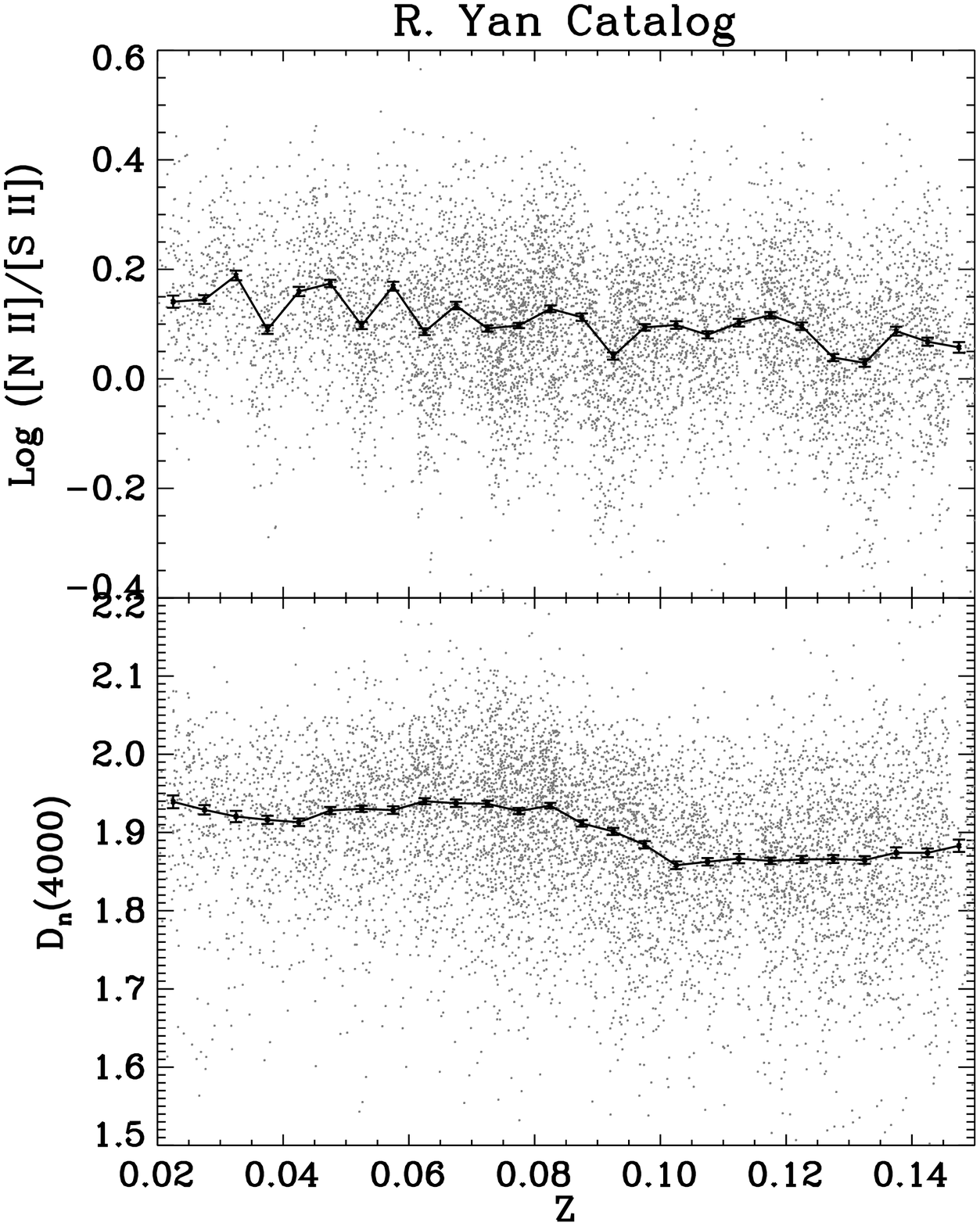}
\includegraphics[totalheight=0.35\textheight]{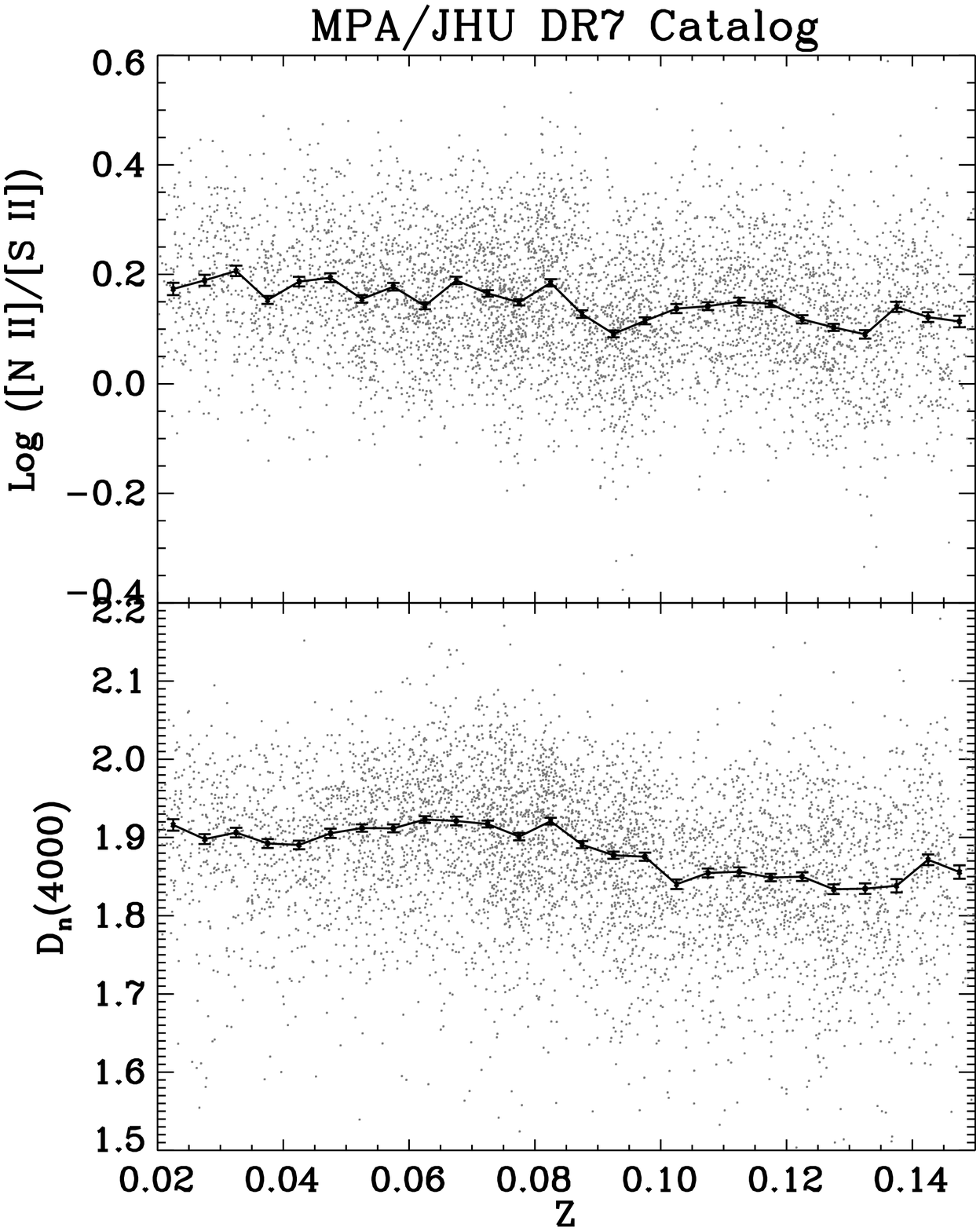}
\caption{Left panels show the distribution of the \niibw/\siiw\ flux ratio (upper panel) and $D_n(4000)$ (lower panel) for LINERs as a function of redshift. The sample is volume-limited with $M_{^{0.1}r} < -20.7$. LINERs are selected using the criteria described by \cite{Kewley06}, except that no criterion involving \oiw\ is required. The dark points with tiny error bars show the median of the distribution. The error bar indicates the error of the median, computed using the formular given by \cite{Beers90}. The wiggly systematic variations with redshift are unphysical. The right panels are similar to the left but using measurements from the MPA/JHU DR7 catalog. The exact measurement methods differ but they all show significant systematic variations with redshift, which are due to a small-scale systematic flux calibration residual.
}
\label{fig:diagnosis_before}
\end{center}
\end{figure*}

Overall, the wavelength-dependent relative flux calibration in SDSS is good to the level of 1\%-2\%. However, this accuracy is still inadequate in certain science applications. 
The problem is illustrated in Figure~\ref{fig:diagnosis_before}. Here, we show the \nii/\sii\ flux ratio and $D_n(4000)$ index for a population of galaxies that have emission lines with ratios characteristic of low-ionization nuclear emission-line regions (LINERs). The sample is volume-limited. Their \nii/\sii\ ratios vary systematically with redshift, at a level nearly comparable to the intrinsic scatter at a single redshift. Similarly, the $D_n(4000)$ index also show unphysical variations with redshift. 

These emission line fluxes and spectral indices are measured with our own software, which is an improved version of the code described by \cite{Yan06}. Similar systematics also exist if we adopt the DR7 catalog produced by the MPA--JHU collaboration\footnote{http://www.mpa-garching.mpg.de/SDSS/} \citep{TremontiHK04}, as shown in the right panels of Figure~\ref{fig:diagnosis_before}. The detailed wiggles differ between our measurements and the MPA/JHU catalog. These differences are mainly due to the difference in the methods used to measure the continuum.\footnote{Our code measures the continuum in the sidebands bracketing the emission line, while the MPA/JHU code measures it using a 200 pixel median smoothing of the emission-line-subtracted continuum.} However, the root cause of the systematic variation with redshift in both cases is due to a small-scale ($\sim10$\AA) systematic flux calibration residual, as we will demonstrate in this paper. 


In this paper, we will establish a new method for improving the flux calibration accuracy by an order of magnitude using coadded red galaxy spectra. Because red galaxies have very uniform spectra and evolve very slowly with redshift, when coadded they provide a high signal-to-noise standard. In addition, because this standard can be found at any redshift, no wavelength will be at disadvantage in the calibration due to features in the standard. We will show that this method can achieve a relative flux calibration accuracy of 0.1\% on scales of a few hundred of Angstroms in SDSS. 


In Section 2, we show how to measure the flux calibration residual in the SDSS using stacked red galaxy spectra, and test the improved spectrophotometry. We discuss potential improvements in Section 3 and summarize in Section 4.

\section{Improved Spectrophotometry}

We employ the SDSS Data Release 7 Main galaxy sample from the New York University Value-Added Galaxy Catalog \citep{BlantonSS05}. The absolute magnitudes are derived using Blanton \& Roweis's (\citeyear{BlantonR07}) {\it k-correct} codes (version 4\_2).

The first step in our calibration procedure is to identify red galaxies with old stellar populations. We first apply two cuts in $^{0.1}(g-r)$ color to select the red sequence galaxies:
\begin{align}
   ^{0.1}(g-r) &> -0.02(M_{^{0.1}r}-5\log h) +0.49 \\
   ^{0.1}(g-r) &< -0.02(M_{^{0.1}r}-5\log h) +0.59 ,
\end{align}
where all magnitudes are in the AB system. The superscript $0.1$ indicates that the filter systems are shifted blueward by a factor of 1.1.

We bin all red galaxies with $0.06<z<0.15$ that satisfy the above criteria and have apparent $r$-band model magnitude brighter than 17.7 into 118 narrow redshift bins. The binsize corresponds to 3 pixels in the SDSS wavelength grid, which is logarithmically spaced ($\Delta\log \lambda = 10^{-4}$). This spacing translates to $\Delta \log (1+z) = 3\times10^{-4}$. We do not need to apply any absolute magnitude limit to our sample, because we will only compare each redshift bin with its neighbors, and the redshift differences between adjacent bins are tiny ($\sim0.0008$). Essentially, we are always comparing samples with the same luminosity distribution. 

Some red-sequence galaxies have star formation but appear red due to dust extinction. To make our standard as stable as possible, we exclude, in each redshift bin, the 20\% of red sequence galaxies that have the lowest $D_n(4000)$ measurement. The final number of galaxies in each redshift bin ranges from 700 to 1300. 

We first de-redden the spectra for galactic extinction according to the dust map of \cite{SchlegelFD98} and \cite{O'Donnell94} extinction curve. Each spectrum is then de-redshifted and smoothed to the same resolution of $350{\rm km s}^{-1}$ by convolving it with a Gaussian kernel with varying width, determined according to the fixed instrumental resolution and the measured velocity dispersion of each galaxy. We normalize each spectrum so that they all have the same median flux between 6010\AA\ and 6100\AA. Finally, we average them in each redshift bin to obtain the stacked spectrum. 
\begin{figure*}
\begin{center}
\includegraphics[totalheight=0.6\textheight,angle=90,viewport=0 10 350 740,clip]{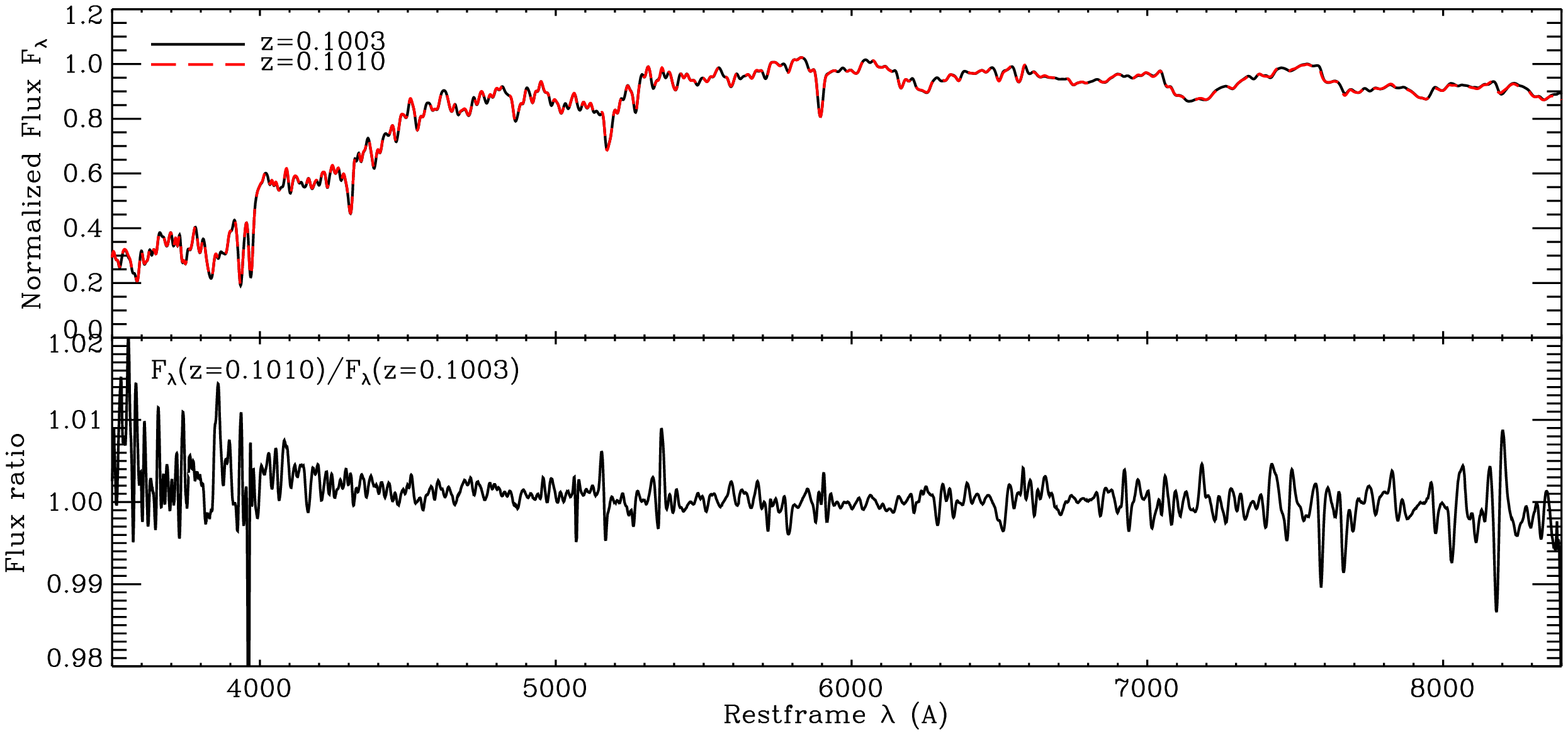}
\caption{Top panel shows a comparison of two stacked spectra of red-sequence galaxies from adjacent narrow redshift bins near $z\sim0.1$ (one solid black curve and one dashed red curve). The two spectra overlap to very high precision. Bottom panel shows the ratio between the two spectra. The tiny differences between the stacked spectra are dominated by spectrophotometric calibration residuals, as the same rest-frame features are redshifted to slightly different wavelengths. The intrinsic differences between the two populations are subdominant as demonstrated below.} 
\label{fig:coadd_diff}
\end{center}
\end{figure*}

The top panel of Figure~\ref{fig:coadd_diff} shows two stacked spectra from a pair of adjacent redshift bins near $z\sim0.1$. There are very tiny, nearly invisible differences between these two coadds. The ratio between the two stacks (bottom panel) varies with wavelength slightly, deviating from 1 by a few percent. The differences have two possible sources. One possibility is intrinsic difference between the two populations, which is a function of rest-frame wavelength. The other possibility is the flux calibration residual, since the same spectral features are redshifted to slightly different wavelengths. The latter is a function of the observed wavelength. We can check which source dominates the difference by comparing the ratios from multiple pairs of stacks in the rest frame and in the observed frame.

\begin{figure*}
\begin{center}
\includegraphics[totalheight=0.75\textheight,angle=90,viewport=0 10 350 740,clip]{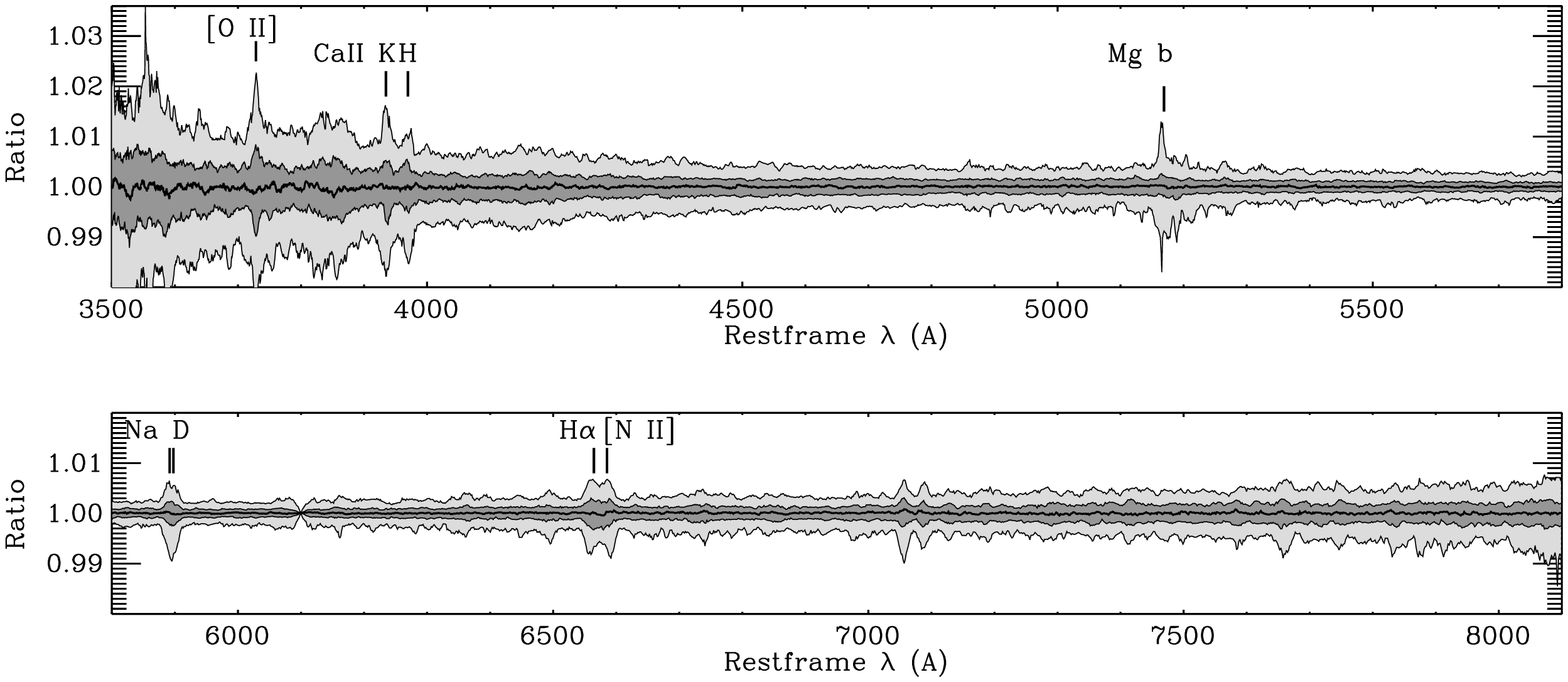}
\caption{Distribution of ratios between pairs of stacked spectra from adjacent redshift bins as a function of {\it rest-frame} wavelength. There are 117 ratio curves from 118 stacked spectra. The thick black line shows their median, and the thinner lines represent the 5, 25, 75, and 95 percentiles at each pixel. The systematic difference among these stacked spectra is tiny, which demonstrates that they can be used as spectroscopic standards to calibrate spectrophotometry. See Figure~\ref{fig:ratio_obsframe}.}
\label{fig:ratio_restframe}
\end{center}
\end{figure*}

In Figure~\ref{fig:ratio_restframe}, we take the ratios from 117 pairs of stacked spectra in adjacent redshift bins, normalize them at 6100\AA, and plot them against the rest-frame wavelength. (We always divide the higher-$z$ bin stack by the lower-$z$ bin stack in each pair.) Lining them up in rest frame smears out the flux calibration difference but highlights the intrinsic population difference. There is very little consistent variation with wavelength among these ratio curves. 
The median ratio has tiny deviations from 1. Over the wavelengths between 3553.8\AA\ and 8072.4\AA\ which are completely covered by all 117 ratio curves, the standard deviation over all pixels in the median curve is 0.00027; only 1.3\% of all pixels have a median ratio deviating from 1 by more than 0.1\%. 


Although the median curve does not show any significant systematic variation, each ratio curve still contains the intrinsic difference between the two redshift bins, which could be due to a tiny mismatch in galaxy properties between the two populations. This can be seen by the larger variation at particular wavelengths, such as \oiiw, \mgbw, \nadw, \hal, \niibw, etc. The intrinsic difference could also give rise to a small smooth deviation on large scales. For example, the ratio curve in the bottom panel of Figure~\ref{fig:coadd_diff} deviates systematically from 1 on the blue side and has a large-scale downward slope. Because these intrinsic differences happen in both positive and negative directions with roughly equal probabilities, they largely cancel out in the median curve. We could also remove the large-scale component by smoothing in the following analysis.

The generally larger variation on the blue end compared to the red end could also be due to the lower flux and thus smaller signal-to-noise ratios in the blue. This also partly explains the larger variation seen around \ion{Ca}{2} H and K lines. 

\begin{figure*}
\begin{center}
\includegraphics[totalheight=0.75\textheight,angle=90,viewport=0 10 300 740,clip]{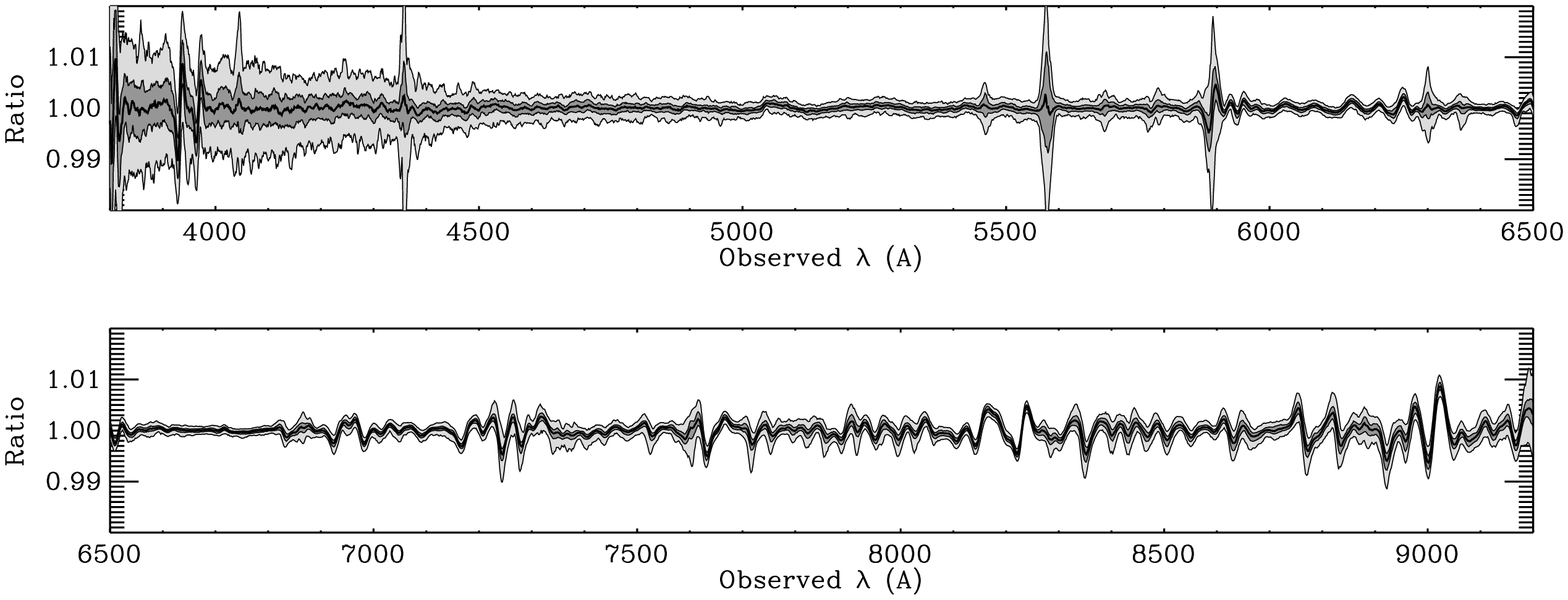}
\caption{Distribution of ratios between pairs of stacked spectra from adjacent redshift bins as a function of {\it observed} wavelength. There are 117 ratio curves from 118 stacked spectra. The thick black line shows their median, while the thinner lines represent the 5, 25, 75, and 95 percentiles at each pixel. Clearly, there is systematic variation with observed wavelength. Comparing this with Figure~\ref{fig:ratio_restframe} shows that the spectrophotometry calibration error dominates the differences between the stacked spectra. Note the tiny scale on the vertical axis. This shows we can achieve high-precision spectrophotometry with stacked spectra of red-sequence galaxies.}
\label{fig:ratio_obsframe}
\end{center}
\end{figure*}

In Figure~\ref{fig:ratio_obsframe}, we plot the same ratio curves against observed wavelength. This smears out the intrinsic population difference but highlights the flux calibration residuals. Here, we divide each ratio curve by its 500 pixel median-smoothed version to remove any large scale component. Any small-scale features due to intrinsic population difference will be smeared out by redshifts and cancelled out among the multiple pairs. Thus, they do not introduce any systematic variation as a function of the observed wavelength. Evidently, the contribution from the flux calibration residual is significant at the level of 1\%, much larger than the median deviation due to population differences ($<0.1\%$). Since our redshift bins are 3 pixels wide, the ratio between two stacked spectra at wavelength $\lambda_k$ ($k$ is the pixel number) is the ratio of the system response function ($R(\lambda)$) between the two 3 pixel wide elements centered on $\lambda_{k+3}$ and $\lambda_k$, respectively. 

Figure~\ref{fig:ratio_restframe} and~\ref{fig:ratio_obsframe} demonstrate that the difference between these stacked spectra can be separated into two main components: a systematic component due to flux calibration error and a noise-like component due to small intrinsic population differences. The latter is only significant blueward of 5000\AA. With many pairs of adjacent redshift bins to average out the ``random'' population differences, the stacked spectra of old red galaxies across small redshift intervals can be used as {\it a spectroscopic standard} to constrain relative system response variations on small scales.




To express the above procedure in mathematical terms, we use $f(\lambda)$ to denote the average spectrum of red galaxies which has been shown to display only tiny systematic variation with redshift. The observed stacked spectrum at redshift $z$ is denoted as $g(\lambda)$. We use $R(\lambda')$ to denote the to-be-constrained system response function in the observed wavelength space ($\lambda'=\lambda(1+z)$). Since we coadded spectra in 3 pixel wide redshift bins, we are effectively using a 3 pixel box smoothed version of $R$, which we denote as $\widetilde R(\lambda')$. 

For two adjacent redshift bins, $z_i$ and $z_{i+1}$, we have
\begin{align}
g_i(\lambda) &= \widetilde R(\lambda(1+z_i)) f_i(\lambda) \\
g_{i+1}(\lambda) &= \widetilde R(\lambda(1+z_{i+1})) f_{i+1}(\lambda)
\end{align}.

We define the ratio, $q_{i,i+1}= g_{i+1}/g_i$. Assuming $f_i=f_{i+1}$, we have
\begin{equation}
\ln \widetilde R(\lambda(1+z_{i+1})) - \ln \widetilde R(\lambda(1+z_i)) = \ln q_{i,i+1}(\lambda)  
\label{eqn:indivi_ratio}
\end{equation}

To rewrite this equation in the observed wavelength space, we set 
\begin{equation}
\lambda_k(1+z_i) =\lambda'_k  , 
\label{eqn:wavek}
\end{equation} 
where $k$ is the pixel index, $k=0,1,\ldots,n-1$, and $n$ is the total number of pixels. Since the wavelength grid is logarithmically spaced, and the spacing between adjacent redshift bins is equivalent to 3 pixels, 
\begin{equation}
\lambda_k(1+z_{i+1})=\lambda'_{k+3}.
\label{eqn:wavek+3}
\end{equation}



We shift each $q_{i,i+1}$ into the observed frame and define 
\begin{equation}
q'_{i,i+1}(\lambda'_k)=q_{i,i+1}(\lambda_k).
\label{eqn:qshift}
\end{equation} 



Substituting Equations (\ref{eqn:wavek})--(\ref{eqn:qshift}) into Equation (\ref{eqn:indivi_ratio}), we have
\begin{equation}
\ln \widetilde R(\lambda'_{k+3}) - \ln \widetilde R(\lambda'_k) = \ln q'_{i,i+1}(\lambda'_k) ,
\end{equation}
where $k= 0,1,\ldots, n-4$; $i = 0,1,\ldots,m-2$; and $m$ is the total number of redshift bins.

Subtracting the smooth component in each $q'_{i,i+1}$ and taking their median at each $\lambda'_k$, we have,
\begin{equation}
\ln \widetilde R(\lambda'_{k+3}) - \ln \widetilde R(\lambda'_k) = \ln \bar q'(\lambda'_k)  
\end{equation}
which we abbreviate as 
\begin{equation}
\ln \widetilde R_{k+3} -\ln \widetilde R_k = \ln q'_k .
\end{equation}

Given this equation array, we would like to solve for $\widetilde R$ using the measured $\bar q'$. With $n$ unknowns but only $n-3$ equations, we are short of three boundary conditions. Inaccurate boundary conditions will introduce oscillations in the solution. Here, we define $\ln \overline R$ to be the 3 pixel box smoothed version of $\ln \widetilde{R}$, i.e., 
\begin{equation}
\ln \overline R_k =  (\ln \widetilde R_{k-1}+ \ln \widetilde R_k + \ln \widetilde R_{k+1})/3 .
\end{equation}
It is easy to show that  
\begin{align}
\ln \overline R_{k+2} -\ln \overline R_{k+1}  = &{\ln \widetilde R_{k+3} - \ln \widetilde R_k \over 3} = \ln \bar q'(\lambda'_k) 
\end{align}

\begin{figure*}
\begin{center}
\includegraphics[totalheight=0.75\textheight,angle=90,viewport=0 0 300 750,clip]{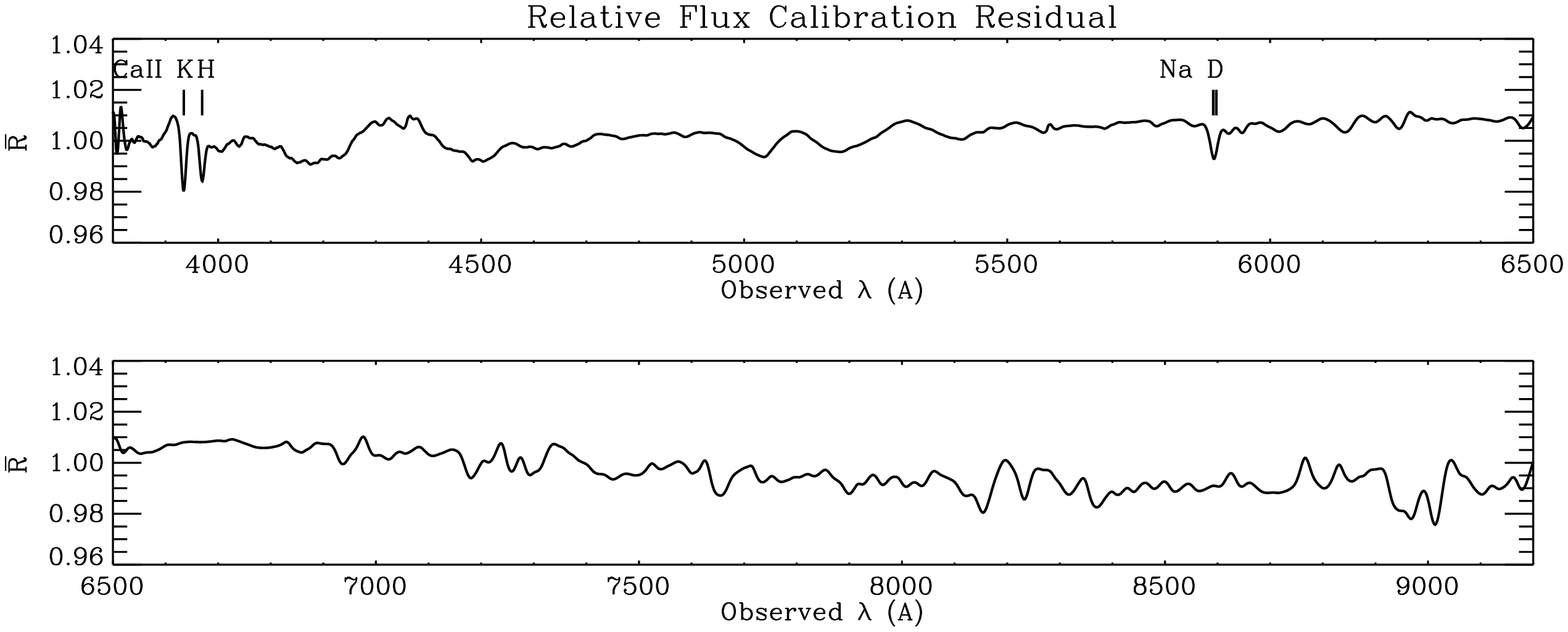}
\caption{Residual system response as a function of observed wavelength. The normalization is arbitrary.}
\label{fig:fluxcalerror}
\end{center}
\end{figure*}

Now it is trivial to solve for $\ln \overline R$ by setting an arbitrary boundary condition. It is arbitrary because we have no constraint on the absolute scale of $\overline R$. 
We then take the exponential and normalize $\overline R$ so that the average between 3800\AA\ and 9200\AA\ is 1. Figure~\ref{fig:fluxcalerror} shows the final spectrophotometry calibration. The correction factor is tabulated in Table~\ref{tab:fluxcalib}. To correct spectra in SDSS, divide the spectra by the vector listed here.

The median $\bar q'$ has a median uncertainty of 0.00009 over all pixels, which is the median uncertainty on the response difference between adjacent 3-pixel-wide elements. Therefore, the final $\overline R$ will have an accumulated error of $\sim0.0009$ over every 300 pixels. Therefore, we have achieved 0.1\% or better accuracy in relative spectrophotometry on wavelength scales shorter than 300--600\AA. 

This uncertainty is of the same order as the uncertainty propagated from the error in the stacked spectra. The latter is limited by the number of galaxies in the stack and the signal-to-noise ratio of each spectrum. Increasing either the signal-to-noise of individual spectra, the number of galaxies in each bin, or the total number of redshift bins can improve the precision of this calibration. The uncertainty is also larger on top of sky lines (atmosphere emission lines, e.g., 5577\AA, 6300\AA) than inbetween sky lines due to the lower signal-to-noise ratio at those wavelengths and the non-Gaussian nature of the sky line subtraction residuals. The larger uncertainties at short wavelengths (see Figure~\ref{fig:ratio_obsframe}) are due to the lower signal-to-noise ratios of red galaxy spectra toward the blue end.

The original flux calibration in SDSS is already accurate to 1\%-2\% level. Now we can remove these few percent residuals to make it accurate to subpercent
level by using red-sequence galaxies as the spectroscopic standard.

\begin{table}[h]
\begin{center}
\caption{Residual System Response ($\overline{R}$) as a Function of Wavelength}
\label{tab:fluxcalib}
\begin{tabular}{c|c}
\hline
\hline
Vacuum Wavelength (\AA) & $\overline{R}$ \\ \hline
3776.5914  &  1.01105 \\
3777.4611  &  1.01877 \\
3778.3310  &  1.02491 \\
3779.2011  &  1.02918 \\
... & ... \\
9238.4689  &  1.00488 \\
9240.5964  &  1.00352 \\
9242.7243  &  1.00060 \\
9244.8528  &  0.99609 \\
9246.9817  &  0.99077 \\
\hline
\end{tabular}
\end{center}
(This table is available in its entirety in machine-readable and Virtual Observatory (VO) forms in the online journal. A portion is shown here for guidance regarding its form and content.)
\end{table}

In this residual correction curve, one can see absorption features at 
3933\AA, 3968\AA, 5890\AA, and 5896\AA, which could be due to \ion{Ca}{2} and 
\ion{Na}{1} absorption in the Milky Way's interstellar medium (ISM), or maybe the 
slightly incorrect F star models around these absorption features. 
There are also features which line up with small scale features in the 
atmospheric transparency curve, such as the water absorption bands around 7150--7400\AA, 8100--8400\AA, and above 8900\AA. 

\begin{figure*}
\begin{center}
\includegraphics[totalheight=0.35\textheight]{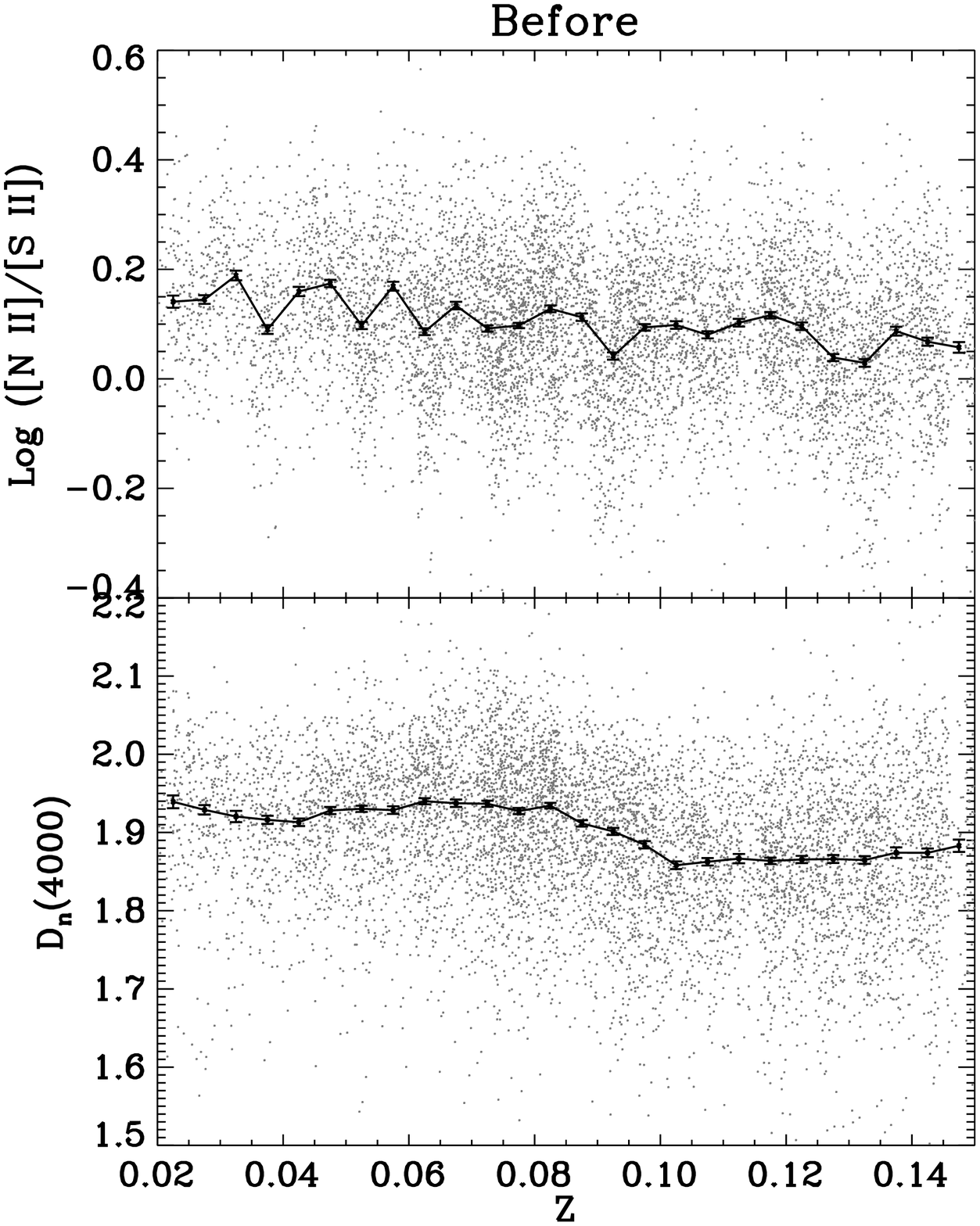}
\includegraphics[totalheight=0.35\textheight]{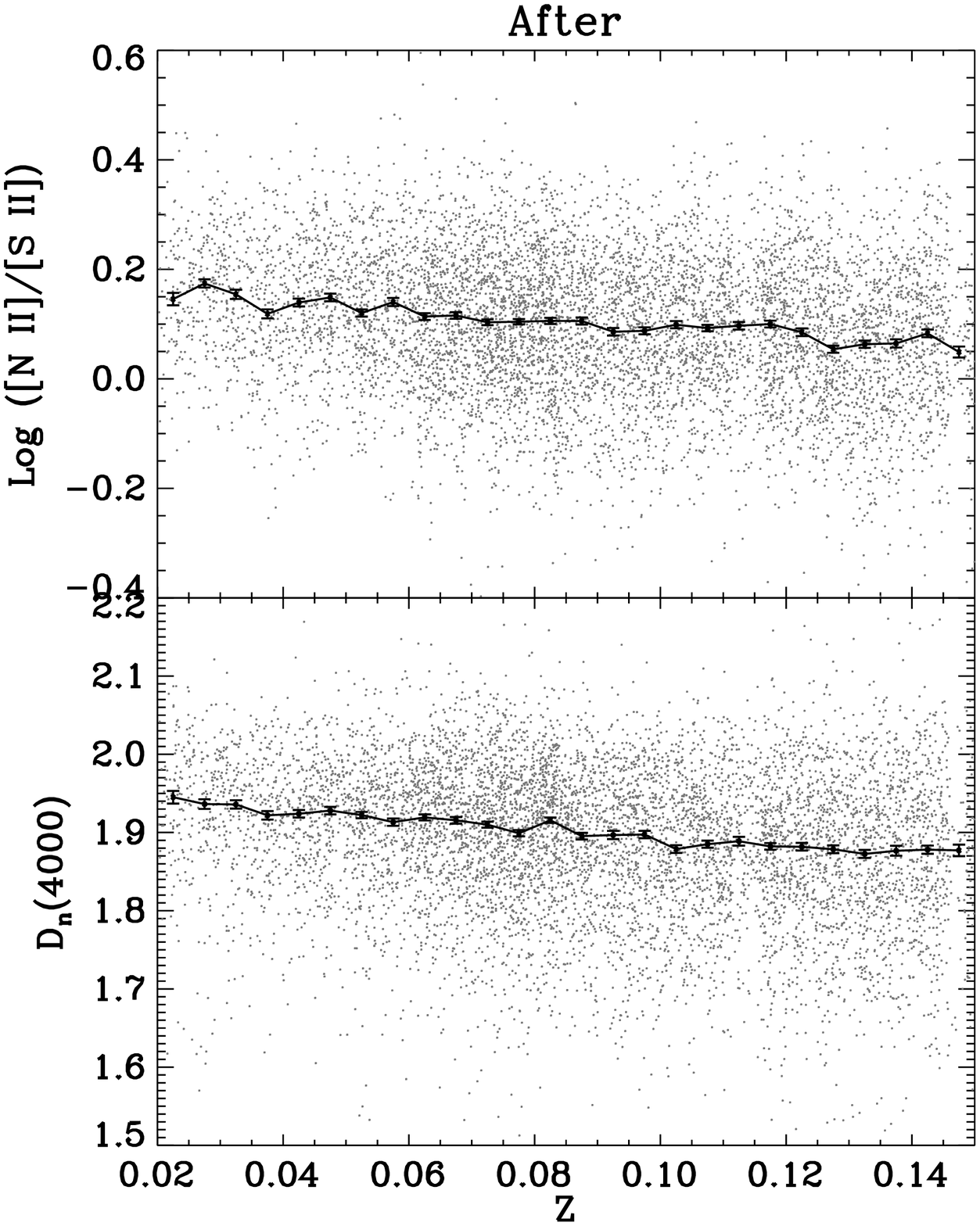}
\caption{Distribution of \niibw/\siiw\ flux ratio (upper panel) and $D_n(4000)$ (lower panel) for LINERs as a function of redshift before (left panel) and after (right panel) applying our flux calibration. The unphysical variations with redshift largely disappear after applying our flux calibration. (The apparent underdensity of points at $z\sim0.114$ is caused by the fact that, at this redshift, the large sky subtraction residual at 5577\AA\ leads to uncertain measurements of \oiiibw, whose significant detection is required for the selection of LINERs.)}
\label{fig:diagnosis_after}
\end{center}
\end{figure*}

To test our improved flux calibration, we applied the correction derived above to the SDSS spectra and remeasured the emission lines for all galaxies in DR7. 
In the right panel of Figure~\ref{fig:diagnosis_after}, we show the \nii/\sii\ ratios and $D_n(4000)$ measurements for LINERs selected in the same manner as in Figure~\ref{fig:diagnosis_before}. The latter is reproduced in the left panel here. These are largely the same galaxies, except that $\sim7\%$ of them differ due to changes in their classifications in the \cite{Kewley06} scheme. This indicates that the impact of the flux calibration on the classification is not negligible. Comparing the two panels, the unphysical systematic trend with redshift is largely gone with the new calibration. The median \nii/\sii\ ratio becomes fairly smooth with redshift. The smooth variation with redshift could result from the aperture effect since the same fiber size corresponds to different physical 
scale at different redshifts. The investigation of these aperture-dependent line ratio variations \citep{YanB11} is precisely what motivated us to improve the small-scale calibration. Though our calibration is still not perfect as shown by the small wiggles left in the right panel of Figure~\ref{fig:diagnosis_after}, it is sufficient for our science project. 

\section{Potential Improvements}
Further improvements of the above calibration are possible along several 
paths. 

First, we could iterate the above calibration process. Since we 
are measuring a smoothed version of $R$, the original $R$ must
have sharper features than what we measure. Thus, folding in a 
first-order correction and iterating would help us measure those sharp 
features more accurately. 

Second, we could use finer redshift bins to get the pixel-to-pixel
variation. However, the advantage of higher pixel resolution needs to be 
weighted against the poorer signal-to-noise in stacked spectra from 
the smaller number of galaxies in each bin. 

Third, as shown in Figure~\ref{fig:ratio_restframe}, certain rest-frame wavelengths in the spectra show larger variation than others, such as \oiiw, \ion{Ca}{2} H and K, \ion{Na}{1} D, \hal, etc. One way to improve our calibration is to exclude these regions (or downweight them) when taking the median (or average) in the observed wavelength space. 

Fourth, the calibration residual has features that might originate from 
three sources: the atmosphere, the instrument, and the ISM. 
These features could have intrinsic variations with different dependencies.
For the atmospheric features, by grouping galaxies according to the observing 
conditions, such as airmass and/or humidity, and doing the calibration 
separately, we could reduce the intrinsic scatter at these wavelengths and 
improve the calibration. 
For features that originate from the telescope and the instrument, 
it could differ between the two spectrographs used in SDSS. It could also vary with the ambient environment during observation or with time 
due to the degrading of mirror coatings. Again, grouping galaxies according 
to the spectrograph, the dates of observation, or ambient environment may 
identify these features
and achieve better corrections. For the absorption in the ISM, galaxies along different lines of sight should have different corrections. 
By grouping red galaxies 
according to their Galactic extinction, we can get better control 
on these wavelengths and may learn something about the ISM at the same 
time. Certain studies may prefer to have the interstellar absorption 
uncorrected. However, we first need to identify robustly the source of
the absorption. The experiment proposed above would also be useful in this 
respect.

Finally, the procedure we have described above does not constrain at all the 
smooth, large-scale variation in the flux calibration. To constrain that, we
could adopt wider redshift bins, a larger step size, and better control for a uniform population of galaxies at all redshifts. In addition, we also need to take into consideration the evolution in the average spectra. It is still possible, since the evolution is a function of the rest-frame wavelength and the spectrophotometry is a function of the observed wavelength. Thus, they can be separated. With a large enough sample, it may be possible to solve for evolution and spectrophotometry simultaneously. 

\section{Summary}

We have demonstrated that the stacked spectrum of carefully selected red-sequence galaxies can be used as a spectrophotometric standard to produce an accurate relative flux calibration for large spectroscopic surveys.
For SDSS, we have achieved the accuracy of $\sim0.1\%$. This method can be applied to surveys like the Baryon Oscillation Spectroscopic Survey (BOSS; \citealt{Eisenstein11}), as long as the survey covers a sufficiently large number of red galaxies. In principle, other types of galaxies could also be utilized provided stacking a large enough sample can produce a reasonably stable spectrum over some wavelengths, which can be verified empirically, as done in Figure~\ref{fig:ratio_restframe}. 

However, this method is not meant to replace the traditional calibration methods using standard stars. To obtain best calibration, we still recommend the practice of observing standard stars simultaneously with science targets, as done in SDSS. They are essential for getting the smooth component in the spectrophotometry and for correcting for those extinction sources that vary on short timescales, such as the molecular absorption in the atmosphere. Our method is meant to be complementary and to improve a reasonably well calibrated spectrophotometry to the accuracy required for certain science applications, such as the study of weak line emission in red galaxies, Ly$\alpha$ forest \citep{Slosar11}, and any studies requiring high accuracy spectral index measurements.



\acknowledgements

I thank Michael Blanton for the encouragement on this work and
for comments on a draft of the paper that helped improve its clarity. I 
also thank John Moustakas, David Hogg, and Guangtun Zhu for helpful 
discussions. The project was supported in part by the NSF Grant AST-0908354
and NASA Grant 08-ADP08-0019.

Funding for the Sloan Digital Sky Survey (SDSS) has been provided by
the Alfred P. Sloan Foundation, the Participating Institutions, the
National Aeronautics and Space Administration, the National Science
Foundation, the U.S. Department of Energy, the Japanese
Monbukagakusho, the Max Planck Society, and the Higher Education Funding
Council for England. The SDSS Web site is http://www.sdss.org/.
The SDSS is managed by the Astrophysical Research Consortium (ARC) for
the Participating Institutions. The Participating Institutions are The
University of Chicago, Fermilab, the Institute for Advanced Study, the
Japan Participation Group, The Johns Hopkins University, Los Alamos
National Laboratory, the Max-Planck-Institute for Astronomy (MPIA),
the Max-Planck-Institute for Astrophysics (MPA), New Mexico State
University, the University of Pittsburgh, Princeton University, the
United States Naval Observatory, and the University of Washington.

\bibliographystyle{apj}
\bibliography{apj-jour,astro_refs}

\end{document}